\documentclass{article}
\usepackage{amssymb}
\usepackage{times}
\newtheorem{thm}{Theorem}[section]
\newtheorem{prop}[thm]{Proposition}

\newcommand{\BZ}{{\mathbb Z}}

\newcommand{\BP}{{\mathbb P}}

\newcommand{\qed}{\hfill$\square$}
\newcommand{\ds}[1]{{\displaystyle #1}}

\newcommand{\dfrac}[2]{{\displaystyle\frac{#1}{#2}}}
\newcommand{\ol}{\overline}
\newcommand{\ul}{\underline}
\def\7b{{\overline 7}}
\def\6u{{\underline 6}}

\newcommand{\og}{\overline {g}}

\def\ll{{\ell}}

\def\df{\underline {f}}

\def\hf{\frac{1}{2}}

\begin{document}
\begin{center}
\textbf{\Large Hypergeometric solutions to the $q$-Painlev\'e
 equations}\\[10mm]
{\large K.~Kajiwara${}^1$, T.~Masuda${}^2$, M.~Noumi${}^2$, 
Y.~Ohta${}^2$ and Y.~Yamada${}^2$}\\[5mm]
${}^1$ Graduate School of Mathematics, Kyushu University,\\
6-10-1 Hakozaki, Fukuoka 812-8581, Japan\\[2mm]
${}^2$ Department of Mathematics, Kobe University,\\
Rokko, Kobe 657-8501, Japan
\end{center}
\begin{abstract}
Hypergeometric solutions to seven $q$-Painlev\'e equations in Sakai's
classification are constructed.  Geometry of plane curves is used to
reduce the $q$-Painlev\'e equations to the three-term recurrence
relations for $q$-hypergeometric functions.
\end{abstract}
\section{Introduction}
It is well-known that the continuous Painlev\'e equations P$_{\rm J}$
(J$=$II,$\ldots$,VI) admit particular solutions expressible in terms of
various hypergeometric functions.
The coalescence cascade of hypergeometric functions, from the Gauss
hypergeometric function to the Airy function, corresponds to that of
Painlev\'e equations, from P$_{\rm VI}$ to P$_{\rm II}$~\cite{GtoP}. 
The similar situation is expected for the discrete Painlev\'e equations.

The discrete Painlev\'e equations and their solutions have been studied
for many years from various view points. 
In particular, Sakai~\cite{Sakai} gave a natural framework of discrete
Painlev\'e equations by means of geometry of rational surfaces.
Among the 22 types in Sakai's classification, there are ten types of
$q$-Painlev\'e equations corresponding to the following degeneration
diagram of type of affine Weyl groups: 
\begin{displaymath}
\begin{array}{l}
E^{(1)}_8\rightarrow E^{(1)}_7\rightarrow E^{(1)}_6\rightarrow D^{(1)}_5
\rightarrow A^{(1)}_4\rightarrow (A_2\!+\!A_1)^{(1)}\rightarrow
(A_1\!+\!A_1^\prime)^{(1)}\rightarrow A_1^{(1)} \rightarrow A_0^{(1)}\\
\phantom{
E^{(1)}_8\rightarrow E^{(1)}_7\rightarrow E^{(1)}_6\rightarrow D^{(1)}_5
\rightarrow A^{(1)}_4\rightarrow (A_2\!+\!A_1)^{(1)}\rightarrow
(A_1\!+\!A_1^\prime)^{(1)}}
\searrow \\
\phantom{
E^{(1)}_8\rightarrow E^{(1)}_7\rightarrow E^{(1)}_6\rightarrow D^{(1)}_5
\rightarrow A^{(1)}_4\rightarrow (A_2\!+\!A_1)^{(1)}\rightarrow
(A_1\!+\!A_1^\prime)^{(1)}
\searrow }\ 
A_1^{(1)\prime} 
\end{array}
\end{displaymath}
In this paper, we construct hypergeometric solutions to the first seven
$q$-Painlev\'e equations. (The last three cases are irrelevant to the
hypergeometric solutions.)

Usually, in order to construct hypergeometric solutions we first look
for special situations in which the discrete Painlev\'e equation is
reducible to the discrete Riccati equation, and linearize it into second
order linear difference equations. 
Then we identify the linear equations with the three-term relations of
appropriate hypergeometric functions. 
Reduction to the discrete Riccati equations has been found for all of
the $q$-Painlev\'e equations~\cite{MSY,RGTT}. 
But so far full step of the above procedure has been carried out only
for $q$-P$_{\rm VI}$ ($D_5^{(1)}$)~\cite{Sakai2}, $q$-P$_{\rm
IV}$~\cite{KNY} and $q$-P$_{\rm III}$~\cite{KK:qp31} ($(A_2+A_1)^{(1)}$)
due to technical difficulty.

In the previous paper~\cite{KMNOY}, we studied the elliptic Painlev\'e
equation, the master equation of all the Painlev\'e and discrete
Painlev\'e equations. 
There we gave an algebraic formulation of $\tau$ functions and a
geometric description of the equation in terms of plane curves.  
We also showed that it admits elliptic hypergeometric function
${}_{10}E_9$ as a hypergeometric solution.
The $q$-Painlev\'e equation with affine Weyl group symmetry of type
$E_8^{(1)}$ can be regarded naturally as a limiting case of the elliptic
Painlev\'e equation.  
The construction of this paper provides explicit hypergeometric solutions
for all the seven types of $q$-Painlev\'e equations.

In Section 2 we give a geometric method to construct hypergeometric
solutions for discrete Painlev\'e equations.
By using this method, we identify the hypergeometric functions appearing
for each $q$-Painlev\'e equation in Section 3.  
In Section 4 we give the list of $q$-Painlev\'e equations and their
explicit hypergeometric solutions.

\section{Decoupling in terms of invariants}
Consider the discrete Riccati equation
\begin{equation}\label{eq:Riccatiabcd}
\overline{x}=\dfrac{ax+b}{cx+d}.
\end{equation}
Here $x=x(t)$ is the unknown variable and $\overline{x}=x(qt)$. 
We will also use the notation $\underline{x}=x(t/q)$.  
In general, the coefficients $a,b,c,d$ also depend on $t$.  
Putting $x=F/G$, the Riccati equation is decoupled into two linear
equations for $F$ and $G$ (the contiguity relations):
\begin{equation}\label{eq:FGcontiguity}
h \overline{F}=aF+bG,\quad
h \overline{G}=cF+dG,
\end{equation}
where $h$ is a decoupling factor.
We then have the following three-term recurrence relations:
\begin{equation}\label{eq:FG3term}
\begin{array}l
h \underline{h} \underline{b}\overline{F}
-\underline{h}(a \underline{b}+b \underline{d})F+
b(\underline{ad}-\underline{bc})\underline{F}=0,\\[2mm]
h \underline{h} \underline{c}\overline{G}
-\underline{h}(d \underline{c}+c \underline{a})G+
c(\underline{ad}-\underline{bc})\underline{G}=0.
\end{array}
\end{equation}
Our task is to solve the Riccati equation (\ref{eq:Riccatiabcd}) through
eqs. (\ref{eq:FGcontiguity}) and (\ref{eq:FG3term}).
For the Riccati equation (\ref{eq:Riccatiabcd}) arising from
$q$-Painlev\'e equations with higher symmetries, the coefficients of
eqs.  (\ref{eq:FG3term}) are polynomials depending on many parameters.
By suitable choice of decoupling factor $h$ and gauge factor $g$ : $F=g
\Phi$, eqs. (\ref{eq:FG3term}) are expected to reduce to some
$q$-hypergeometric equations which typically take the form 
\begin{equation}
A\overline{\Phi}+(B-A-C) \Phi+C\underline{\Phi}=0.
\end{equation}
Here the coefficients $A,B,C$ are of compact factorized form, but,
$B-A-C$ is not. 
Accordingly, the second coefficients in eqs.(\ref{eq:FG3term}) consist
of huge number of terms (more than hundred for $E^{(1)}_8$).  
This is the main technical difficulty to manipulate these equations.
Our basic strategy to overcome this difficulty is to express these
coefficients in terms of invariants such as determinants.  
To do this, the geometric formulation of the discrete Painlev\'e
equations developed in the previous paper~\cite{KMNOY} is useful.
\par\medskip

Consider a configuration of nine points $P_1,\ldots,P_9$ in
$\mathbb{P}^2$.  
We denote by $C_0$ the unique cubic curve passing through them. 
In the context of discrete Painlev\'e equations, the nine points
$P_1,\ldots,P_9$ play the role of parameters for the difference
equations; some of them may be regarded as independent variables. 
An additional generic point $P_{10}$ is regarded as the dependent
variable. 
The commuting family of the time evolutions $T_{ij}$, the translation
associated with a pair of points $(P_i,P_j)$ ($i,j=1,\ldots,9; i\ne j$),
is described as follows.
Let us take $T_{89}$ as an example. 
Under the translation $T_{89}$, the points $P_i$ ($i \neq 8,9,10$) are
invariant and the new points $\overline{P}_8$ and $\overline{P}_9$ are
determined so that
\begin{equation}\label{P8move}
P_8+P_9=\overline{P}_8+\overline{P}_9, \quad
P_1+\cdots+P_7+P_8+\overline{P}_9=0,
\end{equation}
with respect to the addition on the cubic $C_0$, where
$\overline{P}_j=T_{89}(P_j)$. 
This means that $\overline{P}_9$ is the additional intersection point of
the pencil (one parameter family) of cubic curves defined by the eight
points $P_i$ ($i \neq 9,10$). 
Using this pencil of cubics, the transformation $T_{89}(P_{10})$ is
geometrically described as follows. 
Consider a cubic curve $C$ passing through the nine points $P_i$ ($i
\neq 9$). 
The new point $\overline{P}_{10}$ is determined by
\begin{equation}
\overline{P}_{10}+\overline{P}_{9}=P_{10}+P_8,
\end{equation}
with respect to the addition on the curve $C$.

\par\medskip
In view of configuration of nine points $P_1,\ldots,P_{9}$, there are
two typical situations where the dynamical system admits reduction to
discrete Riccati equations.
\par\smallskip\noindent 
\quad(1) The case where three points are collinear. 
\newline
\quad(2) The case where a point is infinitely near to another. 
\par\smallskip\noindent 
For each case we construct below the corresponding Riccati equation and
its linearization.

\paragraph{Case (1):}
This case is further divided into two types depending on the choice of
the translation $T_{ab}$ and the three points $P_l, P_m, P_n$ lying on a
line $\ell$. 
Namely, (1a) $\{a,b\}\cap \{l,m,n\}=\phi$ and (1b) $\{a,b\} \subset
\{l,m,n\}$. 
In both cases, if $P_{10}\in \ell$ then
$\overline{P}_{10}=T_{ab}(P_{10}) \in \ll$ and the motion of $P_{10}$ is
described by a discrete Riccati equation on the line $\ell$.

{\bf (1a)}: 
Consider the case when the three points $P_5,P_6,P_7$ are on a line
$\ll$ and the translation is $T_{89}$ as an example. 
In such a case, $P_{10} \in \ll \Rightarrow \overline{P}_{10} \in \ll$
follows from the fact that the curve $C$ is decomposed into the line
$\ll$ (passing through $P_5, P_6, P_7, \overline{P}_{10}$) and the conic
$C_2$ (passing through $P_1, P_2, P_3, P_4, P_8, \overline{P}_9$).

\begin{prop}\label{prop:Case1a}
The time evolution of the point $P_{10}$ under $T_{89}$ is determined by
the linear equation
\begin{equation}\label{3term-f}
\begin{array}{ll}
\dfrac{d_{jk9}d_{ki8}d_{ij\overline{9}}d_{568}}{d_{i8\overline{9}}}
\left(\lambda_{\{ijk\}}
\dfrac{d_{ik\overline{9}}}{d_{ik8}}d_{jk\overline{10}}-d_{jk10}\right)
\\[10pt]
+
\dfrac{d_{jk8}d_{ki9}d_{ij\underline{8}}d_{569}}{d_{i9\underline{8}}}
\left(\mu_{\{ijk\}}
\dfrac{d_{ik\underline{8}}}{d_{ik9}}d_{jk\underline{10}}-d_{jk10}\right)
=d_{ijk}d_{k89}d_{56j}d_{jk10},
\end{array}
\end{equation}
where $\{ijk\}\subset\{1234\}$ and $d_{abc}=\det[P_a,P_b,P_c]$. 
The gauge factors $\lambda_{\{ijk\}}$ and $\mu_{\{ijk\}}$ can be chosen
as follows:
\begin{equation}
\begin{array}{ll}
\lambda_{\{123\}}=1,\quad
&\lambda_{\{124\}}=\dfrac{(14)}{(13)}=\dfrac{(24)}{(23)},\\[10pt]
\lambda_{\{134\}}=\dfrac{(14)}{(12)}=\dfrac{(34)}{(32)},
\quad 
&\lambda_{\{234\}}=\dfrac{(24)}{(21)}=\dfrac{(34)}{(31)},\\[10pt]
(ij)=\dfrac{d_{ij8}}{d_{ij\overline{9}}},
&\mu_{I}=\lambda_{I}\big|_{8\to 9,\overline{9}\to\underline{8}}.  
\end{array}
\end{equation}
\end{prop}

[Proof] When $\{ijk\}=\{123\}$, eq.(\ref{3term-f}) is reformulation
of~\cite{KMNOY}, Proposition 4.2 in terms of determinants. 
The same proof can be applied for other choice of
$\{ijk\}\subset\{1234\}$. 
The only point we should take care is the gauge factors $\lambda_{I},
\mu_{I}$ that are symmetric with respect to the indices $I=\{ijk\}$. 
By choosing the relative normalization of homogeneous coordinates of
$\overline{P}_{10}$, $P_{10}$ and $\underline{P}_{10}$, one can put
$\lambda_{\{123\}}=\mu_{\{123\}}=1$. 
Then we should determine the other factors $\lambda_{I}$ and $\mu_{I}$,
$I=\{124\}, \{134\}, \{234\}$ consistently. 
To do this, let us consider eqs.(\ref{3term-f}) for $(i,j,k)=(1,2,3)$
and $(i,j,k)=(4,2,3)$, both with the same unknown variable
$d_{2,3,10}$. 
Comparing the corresponding coefficients, we have
\begin{equation}
c_{\{123\}}\lambda_{\{123\}}\dfrac{d_{13\overline{9}}}{d_{138}}=
c_{\{423\}}\lambda_{\{423\}}\dfrac{d_{43\overline{9}}}{d_{438}}, \quad
c_{ijk}=\dfrac{d_{jk9}d_{ki8}d_{ij\overline{9}}d_{568}}
{d_{i8\overline{9}}d_{ijk}d_{k89}d_{56j}}.
\end{equation}
Using the fact that $P_1, P_2, P_3, P_4, P_8$ and $\overline{P}_9$ are
on the conic $C_2$, we have
\begin{equation}\label{coniccond}
\dfrac{c_{123}}{c_{423}}=
\dfrac{d_{381}d_{2\overline{9}1}d_{8\overline{9}4}d_{234}}
{d_{384}d_{2\overline{9}4}d_{8\overline{9}1}d_{231}}=1.
\end{equation}
Hence,
\begin{equation}
\lambda_{\{423\}}=
\lambda_{\{123\}}\dfrac{d_{13\overline{9}}}{d_{138}}
\dfrac{d_{438}}{d_{43\overline{9}}}=\dfrac{(34)}{(31)}=\dfrac{(24)}{(21)}, 
\end{equation}
where the last equality also follows from the conic condition like
eq.(\ref{coniccond}). 
Other factors $\lambda_I$, $\mu_I$ can be determined similarly.\qed

{\bf (1b)}:
Consider next the case when the three points $P_7,P_8,P_9$ are on a line
$\ll$ and translation is $T_{89}$ for instance. 
Then $\overline{P}_{10}\in\ll$ whenever $P_{10}\in\ll$. 
This fact follows from
\begin{equation}
\begin{array}l
P_8+P_{9}=-P_7=\overline{P}_8+\overline{P}_{9},\\
P_8+P_{10}=-P_7=\overline{P}_9+\overline{P}_{10}.
\end{array}
\end{equation}

For a point $P$ with homogeneous coordinates $(x:y:z)$, we set
\begin{equation}
m(P)=[x^3,x^2y,x^2z,xy^2,xyz,xz^2,y^3,y^2z,yz^2,z^3]^t. 
\end{equation}
Then $\overline{P}_{10}$ is determined as a point on $\ell$ such that
\begin{equation}\label{eq:detmP}
X=\det[m(P_1),\ldots,m(P_8),m(P_{10}),m(\overline{P}_{10})]=0. 
\end{equation}

We parameterize the points $P_{10},\overline{P}_{10}\in\ell$ by
setting
\begin{equation}\label{10para}
\begin{array}{l}
P_{10}=d_{179}P_{8} \,v - d_{178} P_{9}=d_{189}P_7+d_{179}(v-1)P_8,\\[6pt]
\overline{P}_{10}=d_{17\overline{9}}\overline{P}_{8}\,\overline{v} - d_{17\overline{8}} \overline{P}_{9}=d_{1\overline{89}}\overline{P}_7+d_{17\overline{9}}(\overline{v}-1)\overline{P}_8.
\end{array}
\end{equation}
The last equality follows from the identity
$d_{ijk}P_l-d_{jkl}P_i+d_{kli}P_j-d_{lij}P_k=0$ and
$d_{789}=d_{7\overline{89}}=0$. 
The coordinate $v$ is chosen so that the three points $P_{10}=P_9, P_7,
P_8$ correspond to $v=0,1,\infty$, respectively (and similar for
$\overline{v}$). 

\begin{prop}\label{prop:Case1b}
The coordinate $\overline{v}$ of $\overline{P}_{10}$ is determined by
the Riccati equation 
\begin{equation}\label{eq:Riccati1b}
\overline{v}=-\dfrac{A_1+A_2\,v}{A_4\,v}.
\end{equation}
In terms of the variable $F$ such that $v=F/\underline{F}$, we obtain
the three-term recurrence relation 
\begin{equation}\label{eq:TTR}
A_4(\overline{F}-F)+A_1(\underline{F}-F)+A_5 F=0.
\end{equation}
The explicit forms of the coefficients $A_i$ are given in the proof.
\end{prop}

[Proof] The determinant $X$ in eq.(\ref{eq:detmP}) is at most cubic with
respect to both the variables $v$ and $\overline{v}$. 
It has trivial zeros at $v=1,\infty$ and $\overline{v}=0,1$
corresponding to $P_{10}=P_7, P_8$ and
$\overline{P}_{10}=\overline{P}_9, P_7$. 
Therefore, $X$ is factorized in the form 
\begin{equation}
X=(v-1)(\overline{v}-1)\overline{v}(A_1+A_2\,v+A_3\,\overline v
+A_4\,v\,\overline{v}).
\end{equation}
First consider $A_3$. 
Comparing the coefficient $v^0 \overline{v}^3$ of $X$, we have 
\begin{equation}
A_3=\det[m(P_1),\ldots,m(P_8),-{d_{178}}^3m(P_9),{d_{17\overline{9}}}^3
m(\overline{P}_8)]=0.
\end{equation}
Next, the coefficient $A_1$ is determined as 
\begin{equation}
A_1=-\dfrac{\partial X}{\partial \overline{v}}\Big|_{v=0,\overline{v}=1}
={d_{178}}^3{d_{1\overline{89}}}^2d_{17\overline{9}}
\det[m(P_1),\ldots,m(P_9),dm(P_7,\overline{P}_8)].
\end{equation}
Here, we have used the relations 
\begin{equation}
\begin{array}l
m(P_{10})=-{d_{178}}^3 m(P_9)+O(v), \\[2mm]
m(\overline{P}_{10})=
{d_{1\overline{89}}}^3 m(P_7)+
(\overline{v}-1){d_{1\overline{89}}}^2d_{17\overline{9}}dm(P_7,\overline{P}_8)
+O((\overline{v}-1)^2),\\[2mm]
dm(P,Q)=\dfrac{d\ }{d\epsilon}m(P+\epsilon\,Q)\big|_{\epsilon=0}. 
\end{array}
\end{equation}
Similarly, the coefficient $A_4$ is given by 
\begin{equation}
\begin{array}{ll}
A_4&=
\dfrac{\partial}{\partial v}
\left({\rm coefficient}\ {\rm of}\ \overline{v}^3\ {\rm in}\ X 
\right)\Big|_{v=1}\\[3mm]
&={d_{17\overline{9}}}^3{d_{189}}^2{d_{179}}
\det[m(P_1),\ldots,m(P_8),dm(P_{7},P_8),m(\overline{P}_8)].
\end{array}
\end{equation}
Finally, for the coefficient $A_2$, it is rather convenient to consider
the combination $A_5=A_1+A_2+A_4$, which is determined by 
\begin{equation}
\begin{array}{ll}
A_5&=
\dfrac{\partial^2 X}{\partial v \partial \overline{v}}\Big|_{v=\overline{v}=1}\\[3mm]
&={d_{1\overline{8}\overline{9}}}^2\, {d_{189}}^2\, d_{17\overline{9}}\,d_{179}
\det[m(P_1),\ldots,m(P_8),dm(P_7,P_8),dm(P_7,\overline{P}_8)].
\end{array}
\end{equation}
The Riccati equation (\ref{eq:Riccati1b}) is derived from $X=0$. 
Since $A_3=0$, it is easily decoupled into eq.(\ref{eq:TTR}).\qed

\paragraph{Case (2)} 
Consider the case where $P_9$ is an infinitely near point of $P_8$
($P_9\to P_8$). 
If $P_{10}$ is also infinitely near to $P_8$, then so is the translation
$\overline{P}_{10}=T_{ab}(P_{10})$ $(a,b \neq 8,9)$. 
The point $\overline{P}_{10}$ is determined by solving the following
system of algebraic equations
\begin{equation}\label{eq:DDD}
\begin{array}{l}
\det[P_a,P_{10},Q]=0,\qquad \det[P_{\overline{b}},P_{\overline{10}},Q]=0,\\
\det[m(P_1),\ldots,\widehat{m(P_b)},\ldots,m(P_8),
dm(P_8,P_9),m(P_{10}),m(Q)]\big|_{\epsilon^3}=0,\\
\det[m(P_1),\ldots,\widehat{m(P_b)},\ldots,m(P_8),
dm(P_8,P_9),m(\overline{P}_{10}),m(Q)]\big|_{\epsilon^3}=0, 
\end{array}
\end{equation} 
including an intermediate infinitely near point $Q$, where
\begin{equation}\label{eq:PPQ}
\begin{array}{l}
P_{10}=P_8+\epsilon \left[\begin{array}{c}0\\u\\v\end{array}\right],\ 
\overline{P}_{10}=P_8+\epsilon 
\left[\begin{array}{l}0 \\\overline{u}\\\overline{v} \end{array}\right],\ 
Q=P_8+\epsilon\left[\begin{array}{l} 0\\r\\s\end{array}\right],
\end{array}
\end{equation}
with an infinitely small parameter $\epsilon$, and $f\big|_{\epsilon^n}$
stands for the coefficient of $\epsilon^n$ in the Taylor expansion of
$f$ at $\epsilon=0$. 
Eq.(\ref{eq:DDD}) can be reduced to a linear relation for the
homogeneous variables $(u:v)$ and $(\overline{u}:\overline{v})$. 
More precisely, there are four solutions to eq.(\ref{eq:DDD}), three of
them are trivial ones: $P_{10}=Q$, $\overline{P}_{10}=Q$ or
$P_{10}=\overline{P}_{10}$, and remaining one gives a linear relation
between $P_{10}$ and $\overline{P}_{10}$ which is in fact the Riccati
equation. 
The variables $(u:v)$ etc. in eq.(\ref{eq:PPQ}) represent the slope of
the line $P_{8}P_{10}$ etc, in a (temporary) coordinate of $\BP^1$ (the
exceptional curve which is the blown up of $P_8$). 
It is convenient to make a change of coordinates $(u:v) \rightarrow
(U:V)=(au+bv:cu+dv)$ in such a way that the lines $P_8P_a$, $P_8P_b$,
$P_8P_9$ correspond to $(0:1),(1:0),(1:1)$, respectively.
Then, in these coordinates, we obtain a three-term recurrence relation for
$F=U/V$ in the form of eq.(\ref{eq:TTR}) with factorized coefficients.

\section{Identification of the hypergeometric functions}
We apply the results in the preceding section to special configurations
of points, and derive the corresponding hypergeometric special solutions
to each of the $q$-Painlev\'e equations in the degeneration diagram
~\cite{Sakai}
\begin{equation}
E^{(1)}_8\rightarrow E^{(1)}_7\rightarrow E^{(1)}_6\rightarrow
D^{(1)}_5\rightarrow A^{(1)}_4\rightarrow (A_2\!+\!A_1)^{(1)}\rightarrow
(A_1\!+\!A_1^\prime)^{(1)}.\label{deg_diagram}
\end{equation}

Let us recall the definition and terminology of the $q$-hypergeometric
series~\cite{Gasper-Rahman}. 
The $q$-hypergeometric series ${}_{r}\varphi_{s}$ is given by,
\begin{equation}
\begin{array}{c}
{}_{r}\varphi_s
\left(
\begin{array}{c}
a_1,\ldots,a_r\\
b_1,\ldots, b_s
\end{array};q,z
\right)=
\ds{\sum_{n=0}^{\infty}}
\dfrac{(a_1;q)_n\cdots (a_r;q)_n}
      {(b_1;q)_n\cdots (b_s;q)_n (q;q)_n}
\left[(-1)^n
      q^{\left(n\atop 2\right)}
\right]^{1+s-r}z^n,\\[12pt]
\qquad(a;q)_n=(1-a)(1-qa)\cdots(1-q^{n-1}a). 
\end{array}
\end{equation}
The $q$-hypergeometric series ${}_{r+1}\varphi_{r}$ is called
\textit{balanced}
\footnote{For ${}_3\varphi_2$ series, it appears that two different
conventions are used in the literature.
This convention is due to~\cite{Gasper-Rahman}, while the series
${}_3\varphi_2\left(\begin{array}{c}a,b,c \\d,e\end{array};q,z\right)$
with $z=de/abc$ is also called ``balanced ${}_3\varphi_2$ series'' in
~\cite{Gupta-Ismail-Masson:1992}.
For ${}_3\varphi_2$ series, we use latter convention without notice.}
if the condition
\begin{equation}
qa_1a_2\cdots a_{r+1}=b_{1}b_2\cdots b_{r},\quad z=q,
\end{equation}
is satisfied, and is called \textit{well-poised} if the condition 
\begin{equation}
qa_1=a_2b_1=\cdots =a_{r+1}b_r,\label{well-poised}
\end{equation}
is satisfied. 
Moreover, it is called \textit{very-well-poised} if it satisfies
\begin{equation}
a_2=qa_1^{\hf},\quad a_3=-qa_1^{\hf},
\end{equation}
in addition to eq.(\ref{well-poised}), and denoted as
${}_{r+1}W_r$:
\begin{equation}
{}_{r+1}W_r(a_1;a_4,\ldots,a_{r+1};q,z)=
{}_{r}\varphi_s\left(
\begin{array}{c}
a_1,qa_1^{\hf},-qa_1^{\hf},a_4\ldots,a_{r+1}\\
a_1^{\hf},-a_1^{\hf},qa_1/a_4,\ldots,qa_1/a_{r+1} 
\end{array};q,z\right).
\end{equation}

The degeneration diagram of $q$-Painlev\'e equations (\ref{deg_diagram})
corresponds to that of $q$-hypergeometric series:
\begin{equation}
\begin{array}{c}
\mbox{balanced} \\
{}_{10}W_9
\end{array}
\rightarrow {}_{8}W_7\rightarrow
\begin{array}{c}
\mbox{balanced} \\
{}_{3}\varphi_2
\end{array}
\rightarrow {}_{2}\varphi_1\rightarrow {}_{1}\varphi_1
\rightarrow
\begin{array}{c}\smallskip
{}_1\varphi_1\left(
\begin{array}{c}
 a\\0
\end{array};q,z\right)\\
{}_1\varphi_1\left(
\begin{array}{c}
0\\b
\end{array};q,z\right)
\end{array}
\rightarrow {}_1\varphi_1\left(
\begin{array}{c} 0\\ -q\end{array};q,z\right). 
\end{equation}

\subsection{Case $E^{(1)}_8$}
In this case we take the configuration of nine points lying on a nodal
cubic curve $C_0$. 
We can parameterize the nine points $P_i=P(u_i)$ as follows:
\begin{equation}
P(u)=\Big(
(1-ubc)(1-u/b)(1-u/c):
(1-uca)(1-u/c)(1-u/a):
(1-uab)(1-u/a)(1-u/b)\Big).
\end{equation}
The function $P(u)$ parameterize a nodal cubic $C_0$ passing through
$(1:0:0), (0:1:0), (0:0:1)$ with a node at $(1:1:1)$. 
Then the determinant $d_{ijk}$ is given by
\begin{equation}\label{eq:dijk}
d_{ijk}=[ij][ik][jk][ijk]d_0,
\end{equation}
where $[ij]=u_i-u_j$, $[ijk]=1-u_iu_ju_k$ and $d_0$ is a constant
independent of $u$.
 
We apply Case (1a) where $P_5,P_6,P_7$ are collinear ($u_5u_6u_7=1$) and
$T=T_{89}$. Putting $(i,j,k)=(1,2,3)$ and substituting the determinants
(\ref{eq:dijk}) in the three-term recurrence relation (\ref{3term-f}),
we obtain the linear equation for $F=d_{2,3,10}$
\begin{equation}\label{eq:E8F}
\begin{array}l
A(l\overline{F}-F)+BF+C(m\underline{F}-F)=0,\\[3mm]
\dfrac{A}{B}=\dfrac{[58][68][138][568][29][2\overline{9}][12\overline{9}][239]}
{[25][26][123][256][89][8\overline{9}][18\overline{9}][389]},\quad
l=\dfrac{[1\overline{9}][3\overline{9}][13\overline{9}]}{[18][38][138]},\\[4mm]
\left(\dfrac{C}{B},m\right)=\left(\dfrac{A}{B},l\right)\Big|_{8
\leftrightarrow 9, \underline{8}\leftrightarrow \overline{9}}~~,
\end{array}
\end{equation}
where $u_{\overline{9}}=qu_9$,$u_{\underline{8}}=qu_8$ and
$qu_1u_2u_3u_4u_8u_9=1$. 
The parameters $u_i$ are transformed by $T_{89}$ as $u_9\mapsto qu_9$,
$u_8\mapsto q^{-1}u_8$ while the other $u_i$ are invariant. 
After the gauge transformation $F=g\,\Phi$ with $\overline{g}=l^{-1}g$,
eq.(\ref{eq:E8F}) is solved by the balanced ${}_{10}W_9$
series~\cite{GM}:
\begin{equation}
\begin{array}{ll}
\smallskip
\Phi&=\Phi(a_0;a_1,\ldots,a_7;q) \\
&={}_{10}W_9(a_0;a_1,\ldots,a_7;q,q)\\
\smallskip
&\displaystyle
\hskip5mm
+\dfrac{(qa_0,a_7/a_0;q)_{\infty}}{(a_0/a_7,qa_7^2/a_0;q)_{\infty}}
\prod_{k=1}^6\dfrac{(a_k,qa_7/a_k;q)_{\infty}}{(qa_0/a_k,a_ka_7/a_0;q)_{\infty}}\\
&\hskip10mm\times\,{}_{10}W_9(a_7^2/a_0;a_1a_7/a_0,\ldots,a_6a_7/a_0,a_7;q,q), 
\end{array}   \label{sol:E8}
\end{equation}
with the balancing condition $q^2a_0^3=a_1a_2\cdots a_7$.  
The parameters $a_i$ ($i=0,1,\ldots,7$) are given by
\begin{equation}
\begin{array}{l}
\smallskip
a_0={u_2^2u_3}/q,\quad
a_1=u_1u_2u_3, \quad
a_2=u_2u_3u_4,\quad
a_3=u_2/u_5,\\
a_4=u_2/u_6,\quad
a_5=u_2u_5u_6,\quad
a_6=u_2u_3u_8,\quad
a_7=u_2u_3u_9.
\end{array}
\end{equation}
When one of the parameters $a_1,a_2,\ldots,a_6$ is $q^{-N}~(N\in\BZ_{\ge
0})$ the second term of eq.(\ref{sol:E8}) vanishes, which can also be
derived as the trigonometric limit of our previous result~\cite{KMNOY}.

\subsection{Case $E^{(1)}_7$}
\begin{figure}
\begin{center}\setlength{\unitlength}{0.5mm}
\begin{picture}(40,40)(0,0)
\put(0,20){\line(1,0){40}}\put(20,20){\circle{40}}
\put(10,20){\circle*{2}}\put(20,20){\circle*{2}}\put(30,20){\circle*{2}}
\put(8,13){1}\put(18,13){2}\put(28,13){6}
\put(10,30){\circle*{2}}\put(20,34){\circle*{2}}\put(30,30){\circle*{2}}
\put(8,33){3}\put(18,37){4}\put(28,33){5}
\put(10,10){\circle*{2}}\put(20,6){\circle*{2}}\put(30,10){\circle*{2}}
\put(8,3){9}\put(18,-1){8}\put(28,3){7}
\end{picture}
\caption{$E^{(1)}_7$}\label{fig:E7}\end{center}
\end{figure}
In this case we consider the configuration of nine points $P_i$ in
$\BP^2$ among which three ($i=1,2,6$) are on a line and six
($i=3,4,5,7,8,9$) are on a conic (Fig.\ref{fig:E7}). 
We parameterize those nine points as follows:
\begin{equation}
P_i=\left\{
\begin{array}{ll}
(-u_i:0:1)&(i=1,2,6)\\[4pt]
(1:u_i:u_i^2)&(i=3,4,5,7,8,9)
\end{array}.
\right.
\end{equation} 
Putting $P_5,P_6,P_7$ to be collinear ($u_5u_6u_7=1$), we again apply
Case (1) with $(i,j,k)=(1,2,3)$. Then we obtain the three-term
recurrence relation (\ref{3term-f}) with respect to $T_{89}$:
\begin{equation}\label{eq:E7F}
\begin{array}l
A(l\overline{F}-F)-BF+C(m\underline{F}-F)=0,\\[3mm]
\dfrac{A}{B}=
\dfrac{qu_9[58][138][568][239]}{u_3u_5[26][89][8\overline{9}][18\overline{9}]},\quad
l=\dfrac{[3\overline{9}][13\overline{9}]}{[38][138]},\\[4mm]
\left(\dfrac{C}{B},m\right)=
\left(\dfrac{A}{B},l\right)
\Big|_{8 \leftrightarrow 9, \underline{8}\leftrightarrow \overline{9}}~~,
\end{array}
\end{equation}
where $F=d_{2,3,10}$ and $q u_1u_2u_3u_4u_8u_9=1$. 
The action of $T=T_{89}$ on the parameters $u_i$ are the same as
$E^{(1)}_8$ case.

By the gauge transformation $F=g\,\Phi$ with
$\overline{g}=l^{-1}\dfrac{[7\overline{9}][258]}{[78][25\overline{9}]}\,g$,
eq.(\ref{eq:E7F}) is solved by the ${}_8W_7$
series~\cite{Ismail-Rahman:AAW},
\begin{equation}
\begin{array}{ll}
\Phi&={}_8W_7(a_0;a_1,\ldots,a_5;q,z),\quad  z=\dfrac{q^2a_0^2}{a_1a_2a_3a_4a_5}, 
\end{array}
\end{equation}
where the parameters $a_i~(i=0,1,\ldots,5)$ are given by
\begin{equation}
\begin{array}{l}
\smallskip
a_0=u_2u_5^2,\quad a_1=u_5/u_8,\quad a_2=u_5/u_9,\\
a_3=qu_5/u_3,\quad a_4=u_5/u_4,\quad a_5=u_2/u_6.
\end{array}
\end{equation}

\subsection{Case $E^{(1)}_6$}
\begin{figure}
\begin{center}\setlength{\unitlength}{0.5mm}
\begin{picture}(80,80)(0,0)
\put(0,20){\line(1,0){80}}\put(10,0){\line(1,2){40}}\put(70,0){\line(-1,2){40}}
\put(30,20){\circle*{2}}\put(40,20){\circle*{2}}\put(50,20){\circle*{2}}
\put(25,30){\circle*{2}}\put(30,40){\circle*{2}}\put(35,50){\circle*{2}}
\put(45,50){\circle*{2}}\put(50,40){\circle*{2}}\put(55,30){\circle*{2}}
\put(28,12){5}\put(38,12){8}\put(48,12){9}
\put(17,30){7}\put(22,40){3}\put(27,50){4}
\put(50,50){1}\put(55,40){2}\put(60,30){6}
\end{picture}
\caption{$E^{(1)}_6$}\label{fig:E6}\end{center}
\end{figure}
In this case the nine points are divided into three groups of three
points that are collinear (Fig.\ref{fig:E6}). 
The parameterization is given as follows:
\begin{equation}
P_i=\left\{
\begin{array}{ll}
(1:-u_i:0)\quad&(i=1,2,6)\\[4pt]
(0:1:-u_i)&(i=3,4,7)\\[4pt]
(-u_i:0:1)&(i=5,8,9)
\end{array}.
\right.
\end{equation}
We again consider the Case (1a) with $P_5,P_6,P_7$ being collinear and
$T=T_{89}$. 
Then the three-term relation (\ref{3term-f}) with $(i,j,k)=(1,2,3)$
implies
\begin{equation}\label{eq:BJE}
\begin{array}{l}
\dfrac{[58][138][239]}{u_1u_3[8\overline{9}]}
\left(\dfrac{[13\overline{9}]}{[138]}
\overline{F}-F\right)
+\dfrac{[59][238][139]}{u_1u_3[\underline{8}9]}
\left(\dfrac{[13\underline{8}]}{[139]}
\underline{F}-F\right)
=-[257][89]F,
\end{array}
\end{equation}
where $F=d_{2,3,10}$. 
This equation is solved by the balanced ${}_3\varphi_2$
series~\cite{Gupta-Ismail-Masson:1992}
\begin{equation}
F={}_3\varphi_2
\left(
\begin{array}{c}
a_1,a_2,a_3\\
b_1,b_2
\end{array};q,\dfrac{b_1b_2}{a_1a_2a_3}
\right), 
\end{equation}
where the parameters are given by
\begin{equation}
\begin{array}{c}
\smallskip
a_1=u_3/u_7,\quad a_2=u_2u_3u_8,\quad a_3=u_2u_3u_9,\\
b_1=u_2u_3u_5,\quad b_2=qu_1u_2u_3^2u_8u_9. 
\end{array}
\end{equation}

\subsection{Case $D^{(1)}_5$}
\begin{figure}
\begin{center}\setlength{\unitlength}{0.5mm}
\begin{picture}(80,80)(0,0)
\put(0,20){\line(1,0){80}}\put(10,0){\line(1,2){40}}\put(70,0){\line(-1,2){40}}
\put(30,20){\circle*{2}}\put(40,20){\circle*{2}}\put(50,20){\circle*{2}}
\put(25,30){\circle*{2}}\put(30,40){\circle*{2}}\put(40,60){\circle*{2}}
\thicklines\put(35,57.5){\line(2,1){10}}\thinlines
\put(50,40){\circle*{2}}\put(55,30){\circle*{2}}
\put(28,12){5}\put(38,12){8}\put(48,12){9}\put(45,57){41}
\put(17,30){7}\put(22,40){3}\put(55,40){2}\put(60,30){6}
\end{picture}
\caption{$D^{(1)}_5$}\label{fig:D5}\end{center}
\end{figure}
This is a limiting case where
$P_1=(\epsilon:1:-u_1\epsilon)\vert_{\epsilon\to 0}$ becomes infinitely
near to $P_4=(0:1:0)$, while the other $P_i$ ($i\neq 1,4$) are the same
as the Case $E^{(1)}_6$ (Fig.\ref{fig:D5}). 
Accordingly, the corresponding linear difference equation is
\begin{equation}\label{eq:LJE}
\begin{array}{l}
\dfrac{u_8[58][239]}{[8\overline{9}]}\left(\dfrac{qu_9}{u_8}\overline{F}-F\right)
+\dfrac{u_9[59][238]}{[\underline{8}9]}\left(\dfrac{qu_8}{u_9}\underline{F}-F\right)
=[257][89]F, 
\end{array}
\end{equation}
with $u_5u_6u_7=1$ and $qu_1u_2u_3u_8u_9=1$. 
This equation is solved by the ${}_{2}\varphi_1$ series~\cite{Sakai2}:
\begin{equation}
F=g\, 
{}_2\varphi_1
\left(
\begin{array}{c}
a_1,a_2\\b_1
\end{array};q,z
\right), \quad \overline{g}=\dfrac{1-u_5/qu_9}{1-u_5/u_8}\,g, 
\end{equation}
where the parameters are given by 
\begin{equation}
a_1=u_2u_3u_8,\quad a_2=u_2u_3u_9,\quad b_1=qu_2u_3u_8u_9/u_5,\quad z=qu_7/u_3. 
\end{equation}

\subsection{Case $A^{(1)}_4$} 
\begin{figure}
\begin{center}\setlength{\unitlength}{0.5mm}
\begin{picture}(80,80)(0,0)
\put(0,20){\line(1,0){80}}\put(10,0){\line(1,2){40}}\put(70,0){\line(-1,2){40}}
\put(30,20){\circle*{2}}\put(40,20){\circle*{2}}\put(50,20){\circle*{2}}
\put(25,30){\circle*{2}}\put(30,40){\circle*{2}}
\thicklines\put(35,70){\line(1,-2){10}}\put(30,55){{\huge $\supset$}}\thinlines
\put(55,30){\circle*{2}}\put(40,60){\circle*{2}}
\put(28,12){1}\put(38,12){2}\put(48,12){3}\put(45,60){789}
\put(17,30){5}\put(22,40){4}\put(60,30){6}
\end{picture}
\caption{$A^{(1)}_4$}\label{fig:A4}\end{center}
\end{figure}
This case is a further degeneration of Case $D^{(1)}_5$:
\begin{equation}
\begin{array}{lll}
P_1=(1:0:1)&
P_2=(a_2:0:1)&
P_3=(a_1a_2:0:1)\\[1mm]
P_4=(0:1:1)&
P_5=(0:1:a_4)&
P_6=(1:-a_3:0)\\[1mm]
P_7=(0:1:0)&
P_8=(\epsilon:1:0)&
P_9=(\epsilon:1:\dfrac{a_0}{a_2}\epsilon^2)
\end{array}, 
\end{equation}
where $\epsilon$ is an infinitesimal parameter and $a_0a_1\cdots a_4=q$.
This configuration contains a sequence of infinitely near points $P_9\to
P_8\to P_7$, while $(P_1,P_2,P_3)$, $(P_4,P_5,P_7)$ and $(P_6,P_7,P_8)$
are collinear (Fig.\ref{fig:A4}).

Consider the case where $P_1,P_5,P_6$ are collinear ($a_3a_4=1$) and the
time evolution is $T=T_{56}$ ($a_3\mapsto a_3/q,~a_4\mapsto a_4
q$). 
This situation corresponds to the Case (1b). 
Applying the Proposition \ref{prop:Case1b}, we obtain the linear
equation,
\begin{equation}
\dfrac{a_2}{a_0}(a_3-q)(\overline{F}-F)
=(\underline{F}-F)+(1-a_2)(1-a_1a_2)F.  \label{eq:A4}
\end{equation}
This equation is solved by 
\begin{equation}
F=g\,
{}_2\varphi_1
\left(
\begin{array}{c}
a_0,a_0a_1\\0
\end{array};q,qa_4
\right), 
\quad \overline{g}=\frac{a_0/a_2}{1-a_3/q}\,g. 
\end{equation}
We note that the above solution can be rewritten in terms of
${}_1\varphi_1$ series by using the relation~\cite{KS}
\begin{equation}
 {}_2\varphi_1\left(\begin{array}{c} a,b\\0 \end{array};q,z\right)
=\frac{(bz;q)_\infty}{(z;q)_\infty}
 {}_1\varphi_1\left(\begin{array}{c} b\\bz \end{array};q,az\right).
\end{equation}

\subsection{Case $(A_2\!+\!A_1)^{(1)}$}
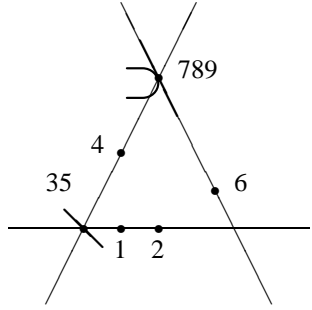
\begin{figure}
\begin{center}\setlength{\unitlength}{0.5mm}
\begin{picture}(80,80)(0,0)
\put(0,20){\line(1,0){80}}\put(10,0){\line(1,2){40}}\put(70,0){\line(-1,2){40}}
\put(30,20){\circle*{2}}\put(40,20){\circle*{2}}
\put(30,40){\circle*{2}}\put(20,20){\circle*{2}}
\thicklines\put(35,70){\line(1,-2){10}}\put(15,25){\line(1,-1){10}}
\put(30,55){{\huge $\supset$}}\thinlines
\put(55,30){\circle*{2}}\put(40,60){\circle*{2}}
\put(10,30){35}\put(45,60){789}
\put(28,12){1}\put(38,12){2}\put(22,40){4}\put(60,30){6}
\end{picture}
\caption{$(A_2\!+\!A_1)^{(1)}$}\label{fig:A2A1}\end{center}
\end{figure}
This case is a further degeneration of Case $A^{(1)}_4$ such that $P_5
\to P_3$ (Fig.\ref{fig:A2A1}):
\begin{equation}
\begin{array}{lll}
P_1=(1:0:1)&
P_2=(a_1:0:1)&
P_3=(0:0:1)\\
P_4=(0:1:1)&
P_5=(-\dfrac{b_1}{a_2}\epsilon:\epsilon:1)&
P_6=(1:-a_2:0)\\
P_7=(0:1:0)&
P_8=(\epsilon:1:0)&
P_9=(\epsilon:1:\dfrac{b_0}{a_1} \epsilon^2)
\end{array}, 
\end{equation}
where $a_0a_1a_2=b_0b_1=q$.  

In this case there are two different kinds of time evolutions, which are
referred to as $q$-P$_{\rm III}$ and $q$-P$_{\rm IV}$, respectively.
The actions on the parameters are given by 
\begin{equation}
\begin{array}{|c|c|ccc|cc|}
\hline
&&a_1&a_2&a_0&b_1&b_0\\
\hline
q\mbox{-P}_{\rm III}&T_{92}&a_1/q&a_2&a_0q&b_1&b_0\\
&T_{56}&a_1&a_2q&a_0/q&b_1&b_0\\
\hline
q\mbox{-P}_{\rm IV}&T_{59}&a_1&a_2&a_0&qb_1&b_0/q\\
\hline
\end{array}~~~. 
\end{equation}
In the case of $q$-P$_{\rm III}$, when $(P_3, P_5, P_6)$ is collinear
($b_1=1$), the linear equation with respect to $T_{92}$ is derived by
taking $(i,j,k)=(1,4,7)$ in eq.(\ref{3term-f}) of Proposition
\ref{prop:Case1a}:
\begin{equation}
a_1(T_{92}(F)-F)+a_1a_2(T_{92}^{-1}(F)-F)+F=0,
\quad F=d_{4,7,10}.
\end{equation}
This equation is solved by Jackson's $q$-Bessel
function~\cite{Gasper-Rahman}
\begin{equation}
F={}_1\varphi_1
\left(
\begin{array}{c}
0\\q/a_2
\end{array};q,a_0
\right).
\end{equation}

For the case of $q$-P$_{\rm IV}$, taking $P_2 \rightarrow P_1$
($a_1=1$), we get the linear equation with respect to $T_{59}$,
\begin{equation}
a_2 T_{59}^{-1}(F)-b_1 T_{59}(F)-(1-b_1)F=0.
\end{equation}
According to the argument of Case (2), this equation is obtained by
taking $F/T_{95}^{-1}(F)$ as the inhomogeneous coordinate of $\BP^1$
such that $0$ and $\infty$ correspond to the lines $P_1P_5$ and
$P_1P_9$, respectively. 
The above equation is solved by 
\begin{equation}
F={}_1\varphi_1
\left(
\begin{array}{c}
a_2\\0
\end{array};q,b_0
\right).
\end{equation}

\subsection{Case $(A_1\!+\!A_1^\prime)^{(1)}$}
\begin{figure}
\begin{center}\setlength{\unitlength}{0.5mm}
\begin{picture}(80,80)(0,0)
\put(0,20){\line(1,0){80}}\put(10,0){\line(1,2){40}}\put(70,0){\line(-1,2){40}}
\put(30,20){\circle*{2}}\put(60,20){\circle*{2}}
\put(30,40){\circle*{2}}\put(20,20){\circle*{2}}
\thicklines\put(35,70){\line(1,-2){10}}\put(15,25){\line(1,-1){10}}
\put(55,15){\line(1,1){10}}\put(30,55){{\huge $\supset$}}\thinlines
\put(40,60){\circle*{2}}\put(10,30){35}\put(45,60){789}\put(28,12){1}
\put(22,40){4}\put(70,25){26}
\end{picture}
\caption{$(A_1\!+\!A_1^\prime)^{(1)}$}\label{fig:A1A1}\end{center}
\end{figure}
This case is obtained from Case $(A_2\!+\!A_1)^{(1)}$ by taking
$P_6 \to P_2$ (Fig.\ref{fig:A1A1}):
\begin{equation}
\begin{array}{lll}
P_1=(1:0:1)&
P_2=(1:0:0)&
P_3=(0:0:1)\\
P_4=(0:1:1)&
P_5=(\dfrac{a_0}{b}\epsilon:\epsilon:1)&
P_6=(1:a_1\epsilon:\epsilon)\\
P_7=(0:1:0)&
P_8=(\epsilon:1:0)&
P_9=(\epsilon:1:-b \epsilon^2)
\end{array}, 
\end{equation}
where $a_0a_1=q$.

When $(P_2,P_4,P_6)$ is collinear ($a_1=1$), we obtain the Riccati
equation
\begin{equation}
T_{95}(y)=\frac{b(1-y)}{y}, \quad P_{10}=(1:y:y), 
\end{equation}
with respect to $T_{95}~(b \mapsto bq)$, which is linearized as 
\begin{equation}
T_{95}(F)-T_{95}^{-1}(F)+q^{-\frac{1}{4}}b^{\frac{1}{2}}F=0, 
\end{equation}
through
\begin{equation}
y=q^{-\frac{1}{4}}b^{\frac{1}{2}}\frac{F}{T_{95}^{-1}(F)}.
\end{equation}
This is solved by a $q$-analogue of the Airy function 
\begin{equation}
F={}_1\varphi_{1}
\left(
\begin{array}{c}
0\\-q^{\frac{1}{2}}
\end{array};q^{\frac{1}{2}},-q^{\frac{1}{4}}b^{\frac{1}{2}}
\right). 
\end{equation}

\section{Hypergeometric solutions to $q$-Painlev\'e equations} 

In the previous section, the relevant hypergeometric functions are
identified for each $q$-Painlev\'e equation. 
However, our choice of the dependent variables and parameters is not
always the same as in the literature. 
In this section we give a list of hypergemetric solutions for the
$q$-Painlev\'e equations in the forms appearing in preceding works.
Full details of construction of the solutions will be given in a forthcoming
paper.

\subsection{Case $E^{(1)}_8$}
\paragraph{$q$-Painlev\'e Equation}\cite{MSY,RGTT,ORG}
\begin{equation}\label{qPe8}
\begin{array}{l}
\dfrac{(\ol{g}\ol{s}t-f)(gst-f)-(\ol{s}^2t^2-1)(s^2t^2-1)}
      {\left(\dfrac{\ol{g}}{\ol{s}t}-f\right)\left(\dfrac{g}{st}-f\right)
      -\left(1-\dfrac{1}{\ol{s}^2t^2}\right)\left(1-\dfrac{1}{s^2t^2}\right)}
=\dfrac{P(f,t,m_1,\ldots,m_7)}{P(f,t^{-1},m_7,\ldots,m_1)},  \\[8mm]
\dfrac{(\ul{f}s\ul{t}-g)(fst-g)-(s^2\ul{t}^2-1)(s^2t^2-1)}
      {\left(\dfrac{\ul{f}}{s\ul{t}}-g\right)\left(\dfrac{f}{st}-g\right)
      -\left(1-\dfrac{1}{s^2\ul{t}^2}\right)\left(1-\dfrac{1}{s^2t^2}\right)}
=\dfrac{P(g,s,m_7,\ldots,m_1)}{P(g,s^{-1},m_1,\ldots,m_7)}, 
\end{array}
\end{equation}
where 
\begin{equation}
\begin{array}{l}
P(f,t,m_1,\ldots,m_7)
=f^4-m_1tf^3+(m_2t^2-3-t^8)f^2\\[1mm]
\hskip30mm +(m_7t^7-m_3t^3+2m_1t)f+(t^8-m_6t^6+m_4t^4-m_2t^2+1), 
\end{array}
\end{equation}
and $m_k$ ($k=1,2,\ldots 7$) are the elementary symmetric functions of
$k$-th degree in $b_i~(i=1,2,\ldots,8)$ with
\begin{equation}
b_1b_2\cdots b_8=1. 
\end{equation}
Moreover,
\begin{equation}
\ol{t}=qt,~t=q^{\hf}s. 
\end{equation}
\paragraph{Constraint on Parameters}\cite{MSY}
\begin{equation}
qb_1b_3b_5b_7=1, \quad b_2b_4b_6b_8=q. \label{qPe8:par}
\end{equation}
\paragraph{Hypergeometric Solution} 
A hypergeometric solution is given by
\begin{equation}
\frac{g-\left(\dfrac{s}{b_1}+\dfrac{b_1}{s}\right)}
     {g-\left(\dfrac{s}{b_8}+\dfrac{b_8}{s}\right)}
=\lambda~\frac{\Phi(q^4a_0;a_1,q^2a_2,\ldots,q^2a_7;q^2)}
              {\Phi(a_0;a_1\ldots,a_7;q^2)},
\end{equation}
where $\Phi$ is the balanced ${}_{10}W_9$ series defined in eq.
(\ref{sol:E8}), and $a_i~(i=0,1,\ldots,7$) and $\lambda$ are given by
\begin{equation}\label{qPe8:a_and_b}
\begin{array}{l}
a_0=\dfrac{1}{qb_1b_2b_8^2},\quad 
a_1=\dfrac{q^2}{b_2b_8t^2}, \quad 
a_2=\dfrac{s^2}{b_2b_8}, \\[4mm]
a_i=\dfrac{b_i}{b_8}\ (i=3,5,7),\quad a_i=\dfrac{b_i}{b_1}\ (i=4,6), 
\end{array}
\end{equation}
and
\begin{equation}
\begin{array}{l}
\lambda=
\dfrac{b_1b_4b_6}{b_8s^2}\,
\dfrac{\left(1-\dfrac{b_4b_6}{b_1b_8}\right)
       \left(1-q^2\dfrac{b_4b_6}{b_1b_8}\right)
       (1-b_3b_5t^2)(1-b_3b_7t^2)(1-b_5b_7t^2)
       {\displaystyle \prod_{i=2,4,6}}\left(1-\dfrac{b_i}{b_1}\right)}
      {\left(1-\dfrac{s^2}{b_1b_8}\right)
       \left(1-\dfrac{q^2s^2}{b_1b_8}\right)
       \left(1-\dfrac{b_4}{b_8}\right)
       \left(1-\dfrac{b_6}{b_8}\right)
       \left(1-\dfrac{q}{b_1b_8s^2}\right)
       {\displaystyle \prod_{i=3,5,7}}\left(1-\dfrac{b_4b_6}{b_1b_i}\right)}, 
\end{array}
\end{equation}
respectively.

\subsection{Case $E^{(1)}_7$}
\paragraph{$q$-Painlev\'e Equation}\cite{MSY,RGTT}
\begin{equation}
\left\{\begin{array}{l}
\dfrac{(\overline{g}f-t\overline{t})(gf-t^2)}{(\overline{g}f-1)(gf-1)}
=\dfrac{(f-b_1t)(f-b_2t)(f-b_3t)(f-b_4t)}{(f-b_5)(f-b_6)(f-b_7)(f-b_8)},
\\[12pt]
\dfrac{(gf-t^2)(g\underline{f}-\underline{t}t)}{(gf-1)(g\underline{f}-1)}
=\dfrac{(g-t/b_1)(g-t/b_2)(g-t/b_3)(g-t/b_4)}
{(g-1/b_5)(g-1/b_6)(g-1/b_7)(g-1/b_8)},
\end{array}\right.
\end{equation}
where
\begin{equation}
\overline{t}=qt,\quad b_1b_2b_3b_4=q,\quad b_5b_6b_7b_8=1 .
\end{equation}
\paragraph{Constraint on Parameters}
\begin{equation}
b_1b_3=b_5b_7. 
\end{equation}
\paragraph{Hypergeometric Solution} A hypergeometric solution is given
by the ${}_8 W_7$ series~\cite{Ismail-Rahman:AAW},
\begin{equation}
z=\dfrac{g-t/b_1}{g-1/b_5}=\dfrac{1-b_3/b_1}
{1-b_3/b_5t}~
\dfrac{
{}_8 W_7\left(
\dfrac{b_1b_8}{b_3b_5};\dfrac{qb_8}{b_5},\dfrac{b_2}{b_3},
\dfrac{b_1t}{b_5},\dfrac{b_1}{b_5t},\dfrac{b_4}{b_3};
q,\dfrac{b_5}{b_6}\right)}
{{}_8 W_7\left(
\dfrac{b_1b_8}{b_3b_5};\dfrac{b_8}{b_5},\dfrac{b_2}{b_3},
\dfrac{b_1t}{b_5},\dfrac{b_1}{b_5t},\dfrac{b_4}{b_3};
q,\dfrac{qb_5}{b_6}\right)}.
\end{equation}
In the terminating case, e.g. $b_4/b_3=q^{-N}$ ($N\in\BZ_{\ge 0}$), the
solution is expressed in terms of the terminating balanced
${}_4\varphi_3$ series (Askey-Wilson polynomials) as 
\begin{equation}
z=\dfrac{1-b_3/b_1}{1-b_3/b_5t}~\frac{
{}_4\varphi_3\left(
\begin{array}{c}
 b_1/b_2,b_1t/b_5,b_1/b_5t,b_4/b_3\\
b_1/b_3,b_1b_4/b_5b_6,b_1b_4/b_5b_8
\end{array};q,q
\right)
}
{
{}_4\varphi_3\left(
\begin{array}{c}
 qb_1/b_2,b_1t/b_5,b_1/b_5t,b_4/b_3\\
qb_1/b_3,b_1b_4/b_5b_6,b_1b_4/b_5b_8
\end{array};q,q
\right)
},
\end{equation}
by using Watson's transformation formula for the terminating ${}_8W_7$
series~\cite{Gasper-Rahman} 
\begin{equation}
\begin{array}{l}
{}_8W_7(a;b,c,d,e,f;q,q^2a^2/bcdef) \\[1mm]
\hskip10pt
=\dfrac{(aq,aq/de,aq/df,aq/ef;q)_\infty}{(aq/d,aq/e,aq/f,aq/def;q)_\infty}
~{}_4\varphi_3\left(
\begin{array}{c}
 aq/bc,d,e,f\\
aq/b,aq/c,def/a
\end{array};q,q
\right).
\end{array}\label{Watson}
\end{equation}

\subsection{Case $E^{(1)}_6$}
\paragraph{$q$-Painlev\'e Equation}\cite{MSY,RGTT}
\begin{equation}
\left\{\begin{array}{l}
(\overline{g}f-1)(gf-1)=t\overline{t}~
\dfrac{(f-b_1)(f-b_2)(f-b_3)(f-b_4)}{(f-b_5t)(f-t/b_5)},
\\[12pt]
(gf-1)(g\underline{f}-1)=t^2~
\dfrac{(g-1/b_1)(g-1/b_2)(g-1/b_3)(g-1/b_4)}{(g-b_6t)(g-t/b_6)},
\end{array}\right.
\end{equation}
where
\begin{equation}
\overline{t}=qt,\quad b_1b_2b_3b_4=1 .
\end{equation}
\paragraph{Constraint on Parameters}
\begin{equation}
b_5b_6=b_1b_2.
\end{equation}
\paragraph{Hypergeometric Solution} 
A hypergeometric solution is given in terms of the balanced
${}_3\varphi_2$ series~\cite{Gupta-Ismail-Masson:1992} as 
\begin{equation}
z=\frac{g-1/b_1}{g-tb_6}=
\dfrac{1-b_3/b_1}{1-b_1b_2b_3t/b_5}
\dfrac{{}_3\varphi_2\left(\begin{array}c
qb_3b_5/t,b_3/b_2,b_1^2b_2b_3\\
qb_3b_5^2/b_2,qb_1b_2b_3^2
\end{array};q,b_5t/b_1\right)}
{{}_3\varphi_2\left(\begin{array}c
b_3b_5/t,b_3/b_2,qb_1^2b_2b_3\\
qb_3b_5^2/b_2,qb_1b_2b_3^2
\end{array};q,b_5t/b_1\right)}.
\end{equation}
In the terminating case, e.g. $b_3/b_2=q^{-N}~(N\in\BZ_{\ge 0})$, the
solution can be rewritten in terms of the terminating ${}_3\varphi_2$
series (big $q$-Jacobi polynomials) as
\begin{equation}
z=\frac{1-b_2/b_1}{1-b_1b_2b_3t/b_5}~
\frac{{}_3\varphi_2
\left(\begin{array}{c}
       b_5t/b_2,b_3/b_2,b_1^2b_2b_3 \\
       qb_3b_5^2/b_2,b_1/b_2
      \end{array};q,q\right)}
     {{}_3\varphi_2
\left(\begin{array}{c}
       b_5t/b_2,b_3/b_2,qb_1^2b_2b_3 \\
       qb_3b_5^2/b_2,qb_1/b_2
      \end{array};q,q\right)},
\end{equation}
by using the formula~\cite{Gasper-Rahman} 
\begin{equation}
\begin{array}{ll}
{}_3\varphi_2
\left(
\begin{array}{c}
a,b,c\\
d,e
\end{array};q,de/abc
\right)
&=\dfrac{(e/b,e/c)_\infty}{(e,e/bc)_\infty}~
{}_3\varphi_2
\left(
\begin{array}{c}
d/a,b,c\\
d,qbc/e
\end{array};q,q
\right)\\[4mm]
&\hskip10pt
+\dfrac{(d/a,b,c,de/bc)_\infty}{(d,e,bc/e,de/abc)_\infty}~
{}_3\varphi_2
\left(
\begin{array}{c}
e/b,e/c,de/abc\\
de/bc,qe/bc
\end{array};q,q
\right).
\end{array}
\end{equation}

\subsection{Case $D^{(1)}_5$}
\paragraph{$q$-Painlev\'e Equation}($q$-Painlev\'e VI equation)
~\cite{Sakai2,Jimbo-Sakai}
\begin{equation}
\left\{\begin{array}{l}
\overline{g}g=\dfrac{(f-a_1t)(f-a_2t)}{(f-a_3)(f-a_4)},\\[12pt]
f\df=\dfrac{(g-b_1t/q)(g-b_2t/q)}{(g-b_3)(g-b_4)},
\end{array}\right.
\end{equation}
where
\begin{equation}
\frac{b_1b_2}{b_3b_4}=q\frac{a_1a_2}{a_3a_4}.
\end{equation}
\paragraph{Constraint on Parameters}
\begin{equation}
\frac{b_1}{b_3}=q\frac{a_1}{a_3},\quad
\frac{b_2}{b_4}=\frac{a_2}{a_4}.
\end{equation}
\paragraph{Hypergeometric Solution}
A hypergeometric solution is given by~\cite{Sakai2},
\begin{equation}
\left\{\begin{array}{l}\medskip
f=a_3~\dfrac{1-a_4b_3/a_3b_4}{1-b_3/b_4}
\dfrac{{}_2\varphi_1\left(\begin{array}c
a_3/a_4,a_2b_4/a_4b_1\\
a_3b_4/a_4b_3
\end{array};q,b_1t/b_3\right)}
{{}_2\varphi_1\left(\begin{array}c
a_3/a_4,qa_2b_4/a_4b_1\\
qa_3b_4/a_4b_3
\end{array};q,b_1t/b_3\right)},\\
g=b_4~\dfrac{1-a_4b_3/a_3b_4}{1-a_4/a_3}
\dfrac{{}_2\varphi_1\left(\begin{array}c
a_3/a_4,qa_2b_4/a_4b_1\\
a_3b_4/a_4b_3
\end{array};q,b_1t/qb_3\right)}
{{}_2\varphi_1\left(\begin{array}c
qa_3/a_4,qa_2b_4/a_4b_1\\
qa_3b_4/a_4b_3
\end{array};q,b_1t/qb_3\right)}.
\end{array}\right.
\end{equation}
In the terminating case, the above solution is expressible in terms of
the little $q$-Jacobi polynomials.

\paragraph{Remark} A class of hypergeometric solutions including the
above solution has been constructed in terms of Casorati determinants in
\cite{Sakai2}.

\subsection{Case $A^{(1)}_4$}
\paragraph{$q$-Painlev\'e Equation}($q$-Painlev\'e V equation)
~\cite{Sakai,RGTT,RGO:qp5}
\begin{equation}
\left\{\begin{array}{l}\medskip
\og g=\dfrac{(f+b_1/t)(f+1/b_1t)}{1+b_3f},\\
f\df=\dfrac{(g+b_2/s)(g+1/b_2s)}{1+g/b_3},
\end{array}\right.
\end{equation}
where
\begin{equation}
\overline{t}=qt,\quad t=q^{\hf}s.
\end{equation}
\paragraph{Constraint on Parameters}
\begin{equation}
b_1b_2b_3^2=q^{-\hf}.
\end{equation}
\paragraph{Hypergeometric Solution}
\begin{equation}
\left\{\begin{array}{l}\medskip
f=\dfrac{1-b_2}{b_2^2b_3}~
\dfrac{{}_2\varphi_1\left(\begin{array}c
q/b_2^2,qb_1^2\\0
\end{array};q,q^{\hf}b_2b_3t\right)}
{{}_2\varphi_1\left(\begin{array}c
1/b_2^2,qb_1^2\\0
\end{array};q,q^{\hf}b_2b_3t\right)},
\\
g=-\dfrac{1}{b_1b_3^2t}~\dfrac{{}_2\varphi_1\left(\begin{array}c
1/b_2^2,b_1^2\\0
\end{array};q,q^{\hf}b_2b_3t\right)}
{{}_2\varphi_1\left(\begin{array}c
1/b_2^2,qb_1^2\\0
\end{array};q,q^{\hf}b_2b_3t\right)}.
\end{array}\right.
\end{equation}
In the terminating case, the solution is expressible in terms of the
alternative $q$-Charlier polynomials or the $q$-Laguerre
polynomials~\cite{KS}. 
As we mentioned in the previous section, the above solution is rewritten
in terms of ${}_1\varphi_1$ series.

\subsection{Case $(A_2\!+\!A_1)^{(1)}$}
\paragraph{$q$-Painlev\'e Equation}($q$-Painlev\'e III equation)
~\cite{RGTT,KNY,KK:qp31,KTGR}
\begin{equation}\label{eqn:qp3}
\left\{\begin{array}{l}\medskip
{\displaystyle \overline{g}gf=b_0~\frac{1+a_0tf}{a_0t+f}},\\
{\displaystyle gf\underline{f}=b_0~\frac{a_1/t+g}{1+ga_1/t}},
\end{array}\right.
\end{equation}
where
\begin{equation}
\overline{t}=qt.
\end{equation}
\paragraph{Constraint on Parameters} 
Eq.(\ref{eqn:qp3}) admits two kinds of specialization which yield
different hypergeometric solutions:
\begin{enumerate}
 \item ${\displaystyle b_0=q}.$
 \item ${\displaystyle a_0a_1=q}.$
\end{enumerate}
\paragraph{Hypergeometric Solution}\cite{RGTT,KNY,KK:qp31}
\begin{enumerate}
 \item 
\begin{equation}
\left\{
\begin{array}{l}\medskip
 g=-\dfrac{a_1}{t}~(1-q^2/a_0^2a_1^2)~\dfrac{
{}_1\varphi_1\left(\begin{array}{l}0 \\q^2a_0^2a_1^2\end{array};q^2,q^2t^2/a_1^2\right)
}
{{}_1\varphi_1\left(\begin{array}{l}0 \\q^4a_0^2a_1^2\end{array};q^2,q^2t^2/a_1^2\right)}, \\
 f=\dfrac{qt}{a_1}~\dfrac{q/a_0a_1}{1-q^2/a_0^2a_1^2}~\dfrac{
{}_1\varphi_1\left(\begin{array}{l}0 \\q^4a_0^2a_1^2\end{array};q^2,q^4t^2/a_1^2\right)
}
{{}_1\varphi_1\left(\begin{array}{l}0 \\q^2a_0^2a_1^2\end{array};q^2,q^2t^2/a_1^2\right)}. 
\end{array}
\right.
\end{equation}
 \item 
\begin{equation}
\left\{
\begin{array}{l}\medskip
 g=\dfrac{b_0}{a_0t}~\dfrac{
{}_1\varphi_1\left(\begin{array}{l}a_0^2t^2 \\0\end{array};q^2,q/b_0\right)
}
{{}_1\varphi_1\left(\begin{array}{l}a_0^2t^2 \\0\end{array};q^2,q^3/b_0\right)}, \\
f=-a_0t~\dfrac{{}_1\varphi_1\left(\begin{array}{l}a_0^2t^2 \\0\end{array};q^2,q^3/b_0\right)}
{{}_1\varphi_1\left(\begin{array}{l}a_0^2t^2 \\0\end{array};q^2,q/b_0\right)}.
\end{array}\right.
\end{equation}
This solution is also expressible in terms of the ${}_1\varphi_1$ series
or a specialization of the ${}_2\varphi_1$ series by using the
formula~\cite{KS} 
\begin{eqnarray}
{}_1\varphi_1\left(\begin{array}{c}z \\ 0 \end{array};q,c\right)
&=&(c;q)_\infty~{}_0\varphi_1\left(\begin{array}{c}- \\ c \end{array};q,cz\right) \\
&=&(c,z;q)_\infty~{}_2\varphi_1\left(\begin{array}{c}0 ,0\\ c \end{array};q,z\right).
\end{eqnarray}
In the terminating case, the solution is expressible in terms of the
Stieltjes-Wigert polynomials~\cite{KS}.
\end{enumerate}
\paragraph{Remark}
Two classes of Casorati determinant solutions which includes the above
       solutions as the simplest cases have been constructed
       in~\cite{KNY,KK:qp31}.

\subsection{Case $(A_1\!+\!A_1^\prime)^{(1)}$}
\paragraph{$q$-Painlev\'e Equation}($q$-Painlev\'e II equation)
~\cite{Sakai,RGTT,RG:coales}
\begin{equation}
(\overline{f}f-1)(f\underline{f}-1)=\frac{at^2f}{f+t},\quad \overline{t}=qt.
\end{equation}
\paragraph{Constraint on Parameter} 
\begin{equation}
 a=q.
\end{equation}
\paragraph{Hypergeometric Solution}
\begin{equation}
f=\frac{{}_1\varphi_{1}\left(
\begin{array}{c}0\\-q\end{array};q,-qt\right)}
{{}_1\varphi_{1}\left(
\begin{array}{c}0\\-q\end{array};q,-t\right)}.
\end{equation}

\end{document}